\documentclass{emulateapj}
\journalinfo{ApJ, in press (v682n1, 20 July 2008)}

\newcommand{\funits}{~$\rm erg~cm^{-2}~s^{-1}$}
\newcommand{\frequnits}{~$\rm s^{-1}$}
\newcommand{\angfrequnits}{~$\rm rad~s^{-1}$}

\newcommand{\velunits}{~$\rm km~s^{-1}$}
\newcommand{\nofluxunits}{~$\rm cm^{2}~s^{-1}$}
\newcommand{\accunits}{~$\rm cm~s^{-2}$}
\newcommand{\elsasser}{Els\"{a}sser}

\def\vec#1{\ensuremath{\mathbf{#1}}}

\newcommand{\revise}{}

\shorttitle{non-WKB Alfv\'en Waves in Multicomponent Winds}
\shortauthors{Li and Li}

\begin{document}
\title{Effects of non-WKB Alfv\'en waves on a multicomponent solar wind with differential ion flow}
\author{Bo Li and Xing Li}
\affil{Institute of Mathematics and Physics, Aberystwyth University,
  SY23 3BZ, UK}
\email{bbl@aber.ac.uk}

\begin{abstract}
We present multicomponent solar wind models
        self-consistently incorporating the contribution from
        dissipationless, monochromatic,
        hydromagnetic (with angular frequencies $\omega$ well below ion gyro-frequencies),
        toroidal Alfv\'en waves, which are coupled to the flow through
        the wave-induced ponderomotive forces.
Protons and alpha particles are treated on an equal footing,
        and the wavelength is not assumed small compared with the spatial scales at which the 
        solar wind parameters vary.
We find that the non-WKB effects are significant, for the fast and slow solar
        wind solutions alike.
In comparison with their non-WKB counterparts the WKB ones are more effective 
        in accelerating the solar wind inside the Alfv\'en point,
        producing significantly enhanced ion fluxes
        and considerably reduced alpha abundance in the inner corona.
Only when $\omega \gtrsim 3.5\times 10^{-3}$ ($1.5\times 10^{-3}$)\angfrequnits\ 
        can the waves in the fast (slow) winds be adequately described by the WKB limit.
Moreover, while the Alfv\'en waves tend to reduce the magnitude of the proton-alpha
       speed difference $|U_{\alpha p}|$ in general, different mechanisms
       operate in two different regimes 
       separated by an $\omega_c\sim \mbox{several}\times 10^{-5}$\angfrequnits.
{\revise This $\omega_c$, defined by Equation~(\ref{eq_def_omegac}), is closely related to the time required by
       a solar wind parcel to traverse an Alfv\'en radius with the speed of center of mass evaluated
       at the Alfv\'en point.}
When $\omega > \omega_c$, the fluctuations are wave-like and tend to accelerate both ion species, thereby
       losing most of their energy by doing work on ion flows;
       whereas when $\omega < \omega_c$, a quasi-static behavior results: the fluctuations may
       act to accelerate the slower 
       flowing ion species but decelerate the faster moving one in a large portion of the computational domain,
       and only a minor fraction of the wave energy flux injected at the base is lost.
The fluctuations with the lowest frequency are no less effective in reducing $|U_{\alpha p}|$ than the WKB waves:
       in the slow solar wind solutions, they may be able to quench 
       a significant $|U_{\alpha p}|$ with base amplitudes as small as $4$\velunits.
The consequences of $\omega_c$ on the velocity fluctuation spectra of protons and alpha particles,
       which are likely to be obtained by future missions like Solar Orbiter and Solar Probe, are discussed.
\end{abstract}

\section{INTRODUCTION}
Alfv\'en waves have both observational and theoretical consequences in
     solar wind studies.
A salient feature in the measured solar wind fluctuations in interplanetary
     space~\citep[see the reviews by, e.g.,][]{TuMarsch_95, Goldstein_etal_95, BrunoCarbone_05},
     Alfv\'en waves may as well account for the Faraday rotation measurements
     inside 10~$R_\odot$~\citep{Hollweg_etal_82}
     and the non-thermal broadening of a number of ultraviolet lines
     measured below $\sim 5$~$R_\odot$
     \citep[e.g.,][]{Banerjee_etal_98, Esser_etal_99}.
It is noteworthy that although the hourly-scale fluctuations seem to be more 
     frequently studied, the fluctuation spectrum measured by Helios
     nevertheless spans a broad frequency range from $10^{-5}$ to $10^{-2}$\frequnits\
     \citep{TuMarsch_95}.
On the theoretical side, it was recognized even before their identification that 
     Alfv\'en waves may provide a ponderomotive force that accelerates 
     the solar wind \citep{Parker_65}.
Furthermore, Alfv\'en waves may be damped, their energy being 
     converted into both thermal and kinetic energies of the flow.
One such damping mechanism that received much attention is based on the idea
     that a turbulent cascade towards high parallel wavenumbers
     transfers the wave energy from the low-frequency hydromagnetic regime
     to the ion cyclotron one, where the energy can be readily
     picked up by ions through cyclotron resonance
     \citep[see the extensive review by][]{HI02}.

Most solar wind models that incorporate the contribution from Alfv\'en waves 
     have been formulated in the WKB limit, 
     where the wavelength is assumed much shorter than
     the spatial scale at which the flow parameters vary.
On the one hand, this considerably simplifies the mathematical treatment.
On the other hand, some measured features of the solar wind fluctuations do posses a number of WKB characteristics.
For instance, the Helios measurements revealed that during hourly-scale Alfv\'enic activities 
     in both the fast and slow wind,
     protons are heavily perturbed, whereas the velocity of alpha particles suffers
     little distortion \citep{Marsch_etal_81, Marsch_etal_82a, Marsch_etal_82b}.
This discrepancy in the responses of different ion species to Alfv\'en waves is
     consistent with the WKB theory considering that
     the proton--alpha-particle differential speed is of the order of the local Alfv\'en speed.
In addition, for the extensively studied hourly-scale fluctuations in the fast solar wind around 1~AU,
     there tends to exist 
     a high correlation between the perturbed magnetic field $\delta \vec{b}$
     and plasma velocity $\delta\vec{v}$
     \citep{BelcherDavis_71}.
As a matter of fact, if examining the normalized cross-helicity spectrum
     such as given by Fig.2-4 in \citet{TuMarsch_95},
     one may find that the high $\delta\vec{v}-\delta\vec{b}$ correlation
     is present across the whole inertial range for fluctuations in the fast solar wind 
     throughout the inner heliosphere explored by Helios.

As has long been recognized~\citep{HO80} (see also~\citet{Goldstein_etal_86}),
     the applicability of the WKB approximation is questionable
     near the coronal base where the solar wind is
     highly inhomogeneous and the phase speed of Alfv\'en waves is high.
There have been several different approaches to assess the finite wavelength effect on
     Alfv\'en waves.
For instance, \citet{Hollweg_73} extended the WKB analysis to higher order
     and found that the finite-wavelength correction may lead to a 
     reduction of the wave force below 10~$R_\odot$ for typical 
     solar wind parameters.
Another approach was adopted by \citet{HO80} who developed a formalism that is valid for
     small-amplitude Alfv\'en waves with arbitrary frequencies.
This formalism has the advantage that outward and inward propagating
     waves can be explicitly separated, the coupling between the two
     can be interpreted as reflection
     (see also \citet{HollwegIsenberg_07}).
It was shown by \citet{CranmerBalle_05} that this formalism is essentially equivalent to the form
     developed by \citet{Velli_93} who expressed the wave transport in terms of
     the \elsasser\ variables, and
     equivalent to the equations derived by
     \citet{ZhouMatthaeus_90} if neglecting the nonlinear
     terms therein.
Note that these nonlinear terms, which are essential for the development of any turbulence, 
     involve the interaction between outward and inward propagating waves.
The adoption of \elsasser\ variables therefore finds application
     in the one-point closure phenomenology that expresses the damping
     of the solar wind turbulence,
     which at least in the case of solar corona is expected to
     cascade the wave energy towards high perpendicular wavenumbers
     \citep[e.g.,][]{Dmitruk_etal_01, CranmerBalle_05, Verdini_etal_05, VerdiniVelli_07}.
The most recent advance following this direction is the construction
     of solar wind models that incorporate the contribution of Alfv\'en waves
     through both the wave pressure gradient force and turbulent heating
     \citep{Cranmer_etal_07}.

Nearly all the non-WKB analyses were conducted in the framework of 
     the single-fluid Magnetohydrodynamics (MHD).
In this case, the need to apportion the acceleration and heating
     among different species is circumvented.
However, for the solar wind, alpha particles
     cannot be deemed minor given their non-negligible abundance
     and the fact that they may flow considerably faster than protons.
In the fast solar wind, the proton-alpha speed difference $U_{\alpha p}=U_\alpha -U_p$
     may be up to $\sim 150$\velunits\ at 0.3~AU 
     where the proton speed $U_p$ is $\sim 600$\velunits~\citep{Marsch_etal_82a}.
Moreover, a substantial $U_{\alpha p}\sim 100$\velunits\
     was also found by Helios in the slow solar wind in the presence
     of Alfv\'enic fluctuations~\citep{Marsch_etal_81}.
Consequently, alpha particles can play an important role in determining the evolution of Alfv\'en waves.
So far the only study that includes alpha particles in 
     a non-WKB analysis for Alfv\'en waves seems to be that by \citet{LiLi07} (hereafter paper I),
     who demonstrated that substantial deviations exist between the WKB expectation
     and computed wave properties even for the relatively high angular frequency
     $\omega=10^{-3}$\angfrequnits.
For the purpose of developing the formalism, paper I examined only the wave propagation
     in a prescribed low-latitude fast solar wind.
However, the computations indicate that the wave acceleration may alter the 
     proton and alpha particle speeds with a wave amplitude as low as 10\velunits\
     imposed at the coronal base.
This is particularly true for the lowest frequency considered, $\omega=10^{-5}$\angfrequnits, 
     in which case the waves may effectively suppress the proton-alpha differential speed
     in interplanetary space.
Therefore, a natural extension to paper I is to construct solar wind models
     that self-consistently
     incorporate the acceleration exerted on ion species by finite wavelength Alfv\'en waves.
The aim of this paper is to present numerical models thus constructed.

The ponderomotive forces by Alfv\'en waves may also play a part in
     determining the Helium abundance $n_{\alpha p} = n_\alpha/n_p$
     in the inner corona through their influence on the ion flow speeds.
It is well established by in situ measurements for regions $r \gtrsim 0.3$~AU that
     the fast solar wind corresponds to a hardly varying $n_{\alpha p}\sim 4.5\%$,
     and that in the slow wind $n_{\alpha p}$ is highly variable, ranging from $0.4\%$ to $10\%$
     \citep[e.g.,][]{McComas_etal_00}.
However, $n_{\alpha p}$ in the corona is subject to some controversy since it has to be indirectly inferred.
Concerning the remote-sensing measurements, while
      evidence exists that $n_{\alpha p}$ may not be enhanced in coronal holes
      relative to its value for typical fast solar winds
      \citep{LamingFeldman_03},
      there also exists evidence suggesting the opposite~\citep{Delaboudiniere_99}.
On the other hand, an enhancement in $n_{\alpha p}$ at altitudes of one or several tenths of
      solar radius above the limb is nearly inherent to all multi-fluid models,
      whether an empirical heating \citep{Hansteen_etal_97, Lie-Svendsen_etal_03}
      or the ion-cyclotron resonance \citep{XLi_03} is applied to generate the solutions.
In some cases the corona may even become Helium dominated \citep[e.g.,][]{Lie-Svendsen_etal_03}.
Note that in some models, the bottom boundary is chosen to be in the mid-chromosphere.
As a result, the set of governing equations has to be extended to include the neutral
      components of Hydrogen and Helium,
      and the fluxes of both elements in the solar wind show a complex dependence
      on the processes that happen in the chromosphere and transition region.
Neither these processes nor the ion-cyclotron resonance will be considered in the present model, however.
Instead, to isolate the non-WKB effects on $n_{\alpha p}$, 
      we will make our model as simple as possible by
      placing the bottom boundary at the coronal base, and heating the ion species by
      an empirical energy flux.

The paper is organized as follows.
In section~\ref{sec_phys_model}, we give a brief overview of the equations
      appropriate for a system consisting of non-WKB Alfv\'en waves and a multi-component 
      solar wind flow.
Some further details on the numerical implementation of the model are 
      given in section~\ref{sec_num_model}.
The numerical solutions are then presented 
      in section~\ref{sec_num_res}.
Finally, section~\ref{sec_conc} summarizes the results, ending with some
      concluding remarks.

\section{PHYSICAL MODEL}
\label{sec_phys_model}

The solar wind model consists of three species, electrons ($e$), protons ($p$),
    and alpha particles ($\alpha$).
Each species $s$ ($s=e, p, \alpha$) is characterized by
    its density $n_{s}$,  velocity $\vec{v}_{s}$,
    mass $m_{s}$, electric charge $e_{s}$, and temperature $T_{s}$.
The electric charge is also measured in units of electron charge $e$, i.e.,
    $e_{s} = Z_{s} e$ with $Z_e \equiv -1$ by definition.
Similarly, the ion mass number $A_k$ follows from the relation $m_k = A_k m_p$ ($k=p, \alpha$). 
The mass density of species ${s}$ is $\rho_{s} = n_{s} m_{s}$,
    and the species partial pressure is $p_{s}= n_{s} k_B T_{s}$, where
    $k_B$ is the Boltzmann constant.
Since the wave frequencies of interest are well below the electron
    plasma frequency,
    the condition of quasi-neutrality is guaranteed,
    $n_e= n_p +Z_\alpha n_\alpha$.
Moreover, quasi-zero current is assumed, $\vec{v}_{e}= (n_p \vec{v}_p+ Z_\alpha n_\alpha \vec{v}_\alpha)/n_e$,
    except when the meridional ion momentum equations are derived.

\subsection{General Description}
\label{sec_model_descrip}
Several simplifying assumptions were made in paper I to facilitate the
    derivation of the governing equations to be used in this study.
For instance, symmetry about the magnetic axis is assumed throughout, i.e., 
    $\partial/\partial\phi\equiv 0$ in the spherical coordinate system 
    ($r, \theta, \phi$).
It is assumed that each species considered may be described by the standard
    five-moment equations and that 
    the electron inertia may be neglected ($m_e\equiv0$).
The time-varying multicomponent solar wind 
    is then seen as a system where the finite wavelength Alfv\'en waves
    are superimposed on an otherwise time-independent flow,
    with the waves and background flow interacting with each other solely through the
    wave-induced ponderomotive forces.
The effect of solar rotation is neglected.
As a result, the unperturbed magnetic field $\vec{B}_P$ has no $\phi$ components.
Let $\hat{\vec{e}}_l$ denote the unit vector along $\vec{B}_P$.
A flux-tube coordinate system can then be defined by the base vectors 
    ($\hat{\vec{e}}_l, \hat{\vec{e}}_N, \hat{\vec{e}}_\phi$)
    where $\hat{\vec{e}}_N=\hat{\vec{e}}_\phi\times\hat{\vec{e}}_l$
    (see Fig.\ref{fig_mf}).
Only purely toroidal Alfv\'en waves are considered, i.e.,
    the magnetic field $\vec{B}$ and species velocities $\vec{v}_s$ 
    ($s=e, p, \alpha$) may be written as
\begin{eqnarray}
 \vec{B}= B_l \hat{\vec{e}}_l + b\hat{\vec{e}}_\phi,
 \vec{v}_s= U_s \hat{\vec{e}}_l + w_s \hat{\vec{e}}_N +u_s\hat{\vec{e}}_\phi ,
\end{eqnarray}
    where the lower-case symbols represent perturbations.

As detailed in paper I, the mathematical manipulation centers on the fact
    that for hydromagnetic waves, 
    $\omega$ is many orders of magnitude lower than
    the ion gyro-frequency $\Omega_k = (Z_k e B_l)/(m_k c)$ ($k=p, \alpha$), where
    $c$ denotes the speed of light.
As a consequence, the ion velocity difference vector has to
     be aligned with the instantaneous magnetic field.
Combined with the assumption of quasi-zero current, this leads to
\begin{eqnarray}
 u_k = u_e + \frac{b}{B_l}\left(U_k-U_e\right)  
	\label{eq_ukandue}
\end{eqnarray}
    where $k=p, \alpha$.
Moreover, the velocity perturbations $w_s$ ($s=e, p, \alpha$) are so small that the
    mass and energy exchange between adjacent magnetic flux tubes may be safely neglected.
One needs to retain $w_s$ only in the simplified ion momentum equations to ensure
    the conservation of total momentum.
In other words, the system of vector equations
     is allowed to be expressed as a force balance condition
     across the meridional magnetic field,
     together with a set of transport equations along it.
For simplicity, in the present study the force balance condition is replaced
     by prescribing a meridional magnetic field to be detailed in section~\ref{sec_num_Bkgrnd_Mf},
     and only the equations governing the time-independent multi-component flow
     and the transport of Alfv\'en waves are considered.
In essence, this is a 1-dimensional model in that there is only 
     one independent spatial variable, $l$, the arclength along a given magnetic line of force
     measured from its footpoint at the solar surface.

\subsection{Equations Governing the Wave Transport}
The equations governing the transport of Alfv\'en waves in a multi-component solar wind
     are {\revise the $\phi$ components of the magnetic induction law
     and the total momentum conservation (Eqs.(12) and (16) in paper I).
Using the total momentum as opposed to individual momenta of ion species is necessary to
     eliminate the terms associated with ion gyro-frequencies.
The analysis of the two equations may then be considerably simplified with the introduction of
     the Fourier amplitudes at a given angular frequency $\omega$, i.e.,}
\begin{eqnarray}
 [b(l, t), u_s(l,t)] = [\tilde{b}(l),\tilde{u}_s(l)] \exp(-i\omega t), 
\label{eq_def_Fourier}
\end{eqnarray}
    where $s=e, p, \alpha$.
{\revise Further introducing two dimensionless variables
\begin{eqnarray}
\xi = \tilde{b}/B_l, \hspace{0.4cm} \eta=\tilde{u}_e/U_A,
\label{eq_def_xi_eta}
\end{eqnarray}
      where $U_A = B_l/\sqrt{4\pi\rho}$ is the Alfv\'en speed determined by the total
      mass density $\rho=\rho_p+\rho_\alpha$,
      paper I yields that} the perturbations obey
\begin{eqnarray}
\left(M_T^2-1\right){\xi}' &=& F_{11} \xi +F_{12} \eta , \label{eq_num_xi}\\
\left(M_T^2-1\right){\eta}' &=& F_{21} \xi +F_{22} \eta . \label{eq_num_eta}
\end{eqnarray}
{\revise By doing so the original partial differential equations are reduced to two ordinary ones, which involve 
      only the directional differentiation along the poloidal magnetic field $\partial/\partial l$, denoted by the prime $'$ for brevity}.
In addition, the combined meridional Alfv\'enic Mach number $M_T$ is defined by
\begin{eqnarray}
M_T^2 = 4\pi (\rho_p U_p^2+\rho_\alpha U_\alpha^2)/B_l^2 .
\label{eq_def_MT} 
\end{eqnarray}
The expressions for the complex-valued coefficients $F_{11}, F_{12}, F_{21}$ and $F_{22}$, given by
      Equations (40a) to (40d) in paper I, involve the
      flow parameters $\rho_k$ and $U_k$ ($k=p, \alpha$), the meridional magnetic field $B_l$,
      as well as $\omega$.
Obviously, Equations~(\ref{eq_num_xi}) and (\ref{eq_num_eta}) possess an apparent singularity 
      at the Alfv\'en point where $M_T=1$, 
      which lies between 1~$R_\odot$ and 1~AU for typical solar winds.

Given a line of force and the background flow parameters, 
     equations~(\ref{eq_num_xi}) and (\ref{eq_num_eta}) may be solved for the
     Fourier amplitudes for the electron velocity
     and the magnetic field perturbations if $\omega$ is also known. 
The ion velocity perturbations are then found from the alignment
    condition~(\ref{eq_ukandue}).
By assumption, the feedback from waves to the solar wind flow
    is through the wave induced acceleration $a_{k}$ ($k=p, \alpha$) given by
\begin{eqnarray}
a_{k} &=& \left<u_{k}^2\right> (\ln R)'   
    -\frac{Z_k}{4\pi n_e m_k}\left <b\frac{\partial b}{\partial l}+b^2 (\ln R)'\right> \nonumber\\
    && -\frac{\left<b X_k\right>}{B_l},
\label{eq_def_wave_acce}
\end{eqnarray}
    where the angular brackets represent the time-average over one wave period,
    which may be readily evaluated using the Fourier amplitudes
    (cf. Eq.(21) in paper I).
In addition, $R=r\sin\theta$ is a geometrical factor evaluated at each point
    along the meridional magnetic field line (see Fig.\ref{fig_mf}).
The variable $X_k$ is defined by $X_k=\Omega_k (w_j-w_k)(Z_j n_j/n_e)$, 
    where $j$ stands for the ion species other than $k$,
    i.e., $j=\alpha$ for $k=p$ and vice versa.

Due to the presence of $w_k$ in $X_k$, one may
    expect that the $N$-component of the ion momentum equations has to
    be solved.
In fact, there is no need to do so because $X_k$ involves only the
    difference $w_j-w_k$ and may be determined immediately
    after $u_k$ and $b$ are found, 
\begin{eqnarray}
 X_k &=& \frac{\partial u_k}{\partial t}
     + U_k \left[\frac{\partial u_k}{\partial l} + u_k\left(\ln R\right)'\right] \nonumber \\
     && -\frac{Z_k}{4\pi n_e m_k}B_l\left[\frac{\partial b}{\partial l} + b\left(\ln R\right)'\right] .
\label{eq_def_Xk}
\end{eqnarray}
The algebraic manipulation leading to this rather interesting behavior was first
     devised by \citet{McKenzie_etal_79} when deriving the expressions for
     the force introduced into the meridional ion momentum equation
     by solar rotation.
\citet{McKenzie_etal_79} noted that this practice is in effect an expansion
     in terms of a small parameter $\Omega_\odot/\Omega_k$,
     where $\Omega_\odot$ is the angular rotation rate of the Sun. 
That the technique also applies when
     hydromagnetic Alfv\'en waves are concerned is not surprising,
     given that the Alfv\'en waves in this study are represented by the azimuthal
     twists and there also exists a small parameter $\omega/\Omega_k$.

Equations~(\ref{eq_num_xi}) and (\ref{eq_num_eta}) are analytically tractable in the WKB limit where
      $\omega$ is sufficiently high to ensure that the background flow parameters vary little
      within one wavelength.
To the lowest order, the WKB approximation yields that
      $\left<b^2\right>$ satisfies the equation
\begin{eqnarray}
 \left<b^2\right> \frac{\rho \left(U_{ph}- U_m\right) U_{ph}^2}{B_l^3} = \mbox{const},
\label{eq_wkb_actioncons}
\end{eqnarray}
      where $U_m=(\rho_p U_p + \rho_\alpha U_\alpha)/\rho$ is the speed of center of mass,
      while the phase speed $U_{ph}$ is given by
      the dispersion relation 
\begin{eqnarray}
\sum_k \rho_k (U_{ph}-U_k)^2 = \frac{B_l^2}{4\pi}, 
\label{eq_wkb_disprel}
\end{eqnarray}
      with $k=p, \alpha$.
Furthermore, the amplitudes of the species velocity and magnetic field fluctuations are related by
\begin{eqnarray}
|u_s| = \left(U_{ph}-U_s\right)\frac{|b|}{B_l},
\label{eq_wkb_eigen} 
\end{eqnarray}
where $s=e, p, \alpha$.
Finally, a compact expression is found for the wave acceleration $a_{k}$,
\begin{eqnarray}
 a_{k} = \left(\frac{U_{ph}^2-U_k^2}{2 B_l^2} \left<b^2\right>\right)'.
\label{eq_wkb_acce}
\end{eqnarray}

\subsection{Equations Governing the Time-independent Solar Wind}
From Paper I, the equation governing the meridional speed of ion species $k$ reads
\begin{eqnarray}
U_{k}U_{k}'
&& = -\frac{p_k'}{n_k m_k} 
      -\frac{Z_k p_e'}{n_e m_k}-\frac{G M_\odot}{r} (\ln r)'   \nonumber \\
&&   + \frac{n_j}{A_k n_e} c_0 \left(U_j-U_k\right)
       + a_{k}, \label{eq_vkl} 
\end{eqnarray}
     where $k = p, \alpha$.
The gravitational constant is denoted by $G$,
    and $M_\odot$ is the mass of the Sun.
In addition, $c_0$ is a coefficient associated with Coulomb frictions, 
    which unsurprisingly also involve the ion species $j$
    other than $k$.
We take the Coulomb logarithm to be $21$ when evaluating $c_0$ via
    the expression given in the appendix of \citet{LiLi06}.
Other equations concern the mass conservation for both ion species $k=p, \alpha$,
    and energy transport for all the species $s=e, p, \alpha$.
For simplicity, the ion heat fluxes are neglected, whereas 
    the Spitzer law for the electron heat flux $q_e$ is assumed,
    $q_e = -\kappa T_e^{5/2}T_e'$,
    where $\kappa = 7.8\times 10^{-7}$ {erg}~{K}$^{-7/2}$~{cm}$^{-1}$~s$^{-1}$
    \citep{Spitzer_62}.

If the wave acceleration $a_{k}$ is a known function of $l$,
     the model equations can 
     be solved for the distributions
     along a given magnetic line of force of
     the densities $n_k$ and meridional speed $U_k$
     of ion species ($k=p, \alpha$) as well as the temperatures $T_s$
     of all species ($s=e, p, \alpha$).
However, $a_{k}$ is not known a priori but rather
     depends on the flow parameters themselves, thereby
     coupling the waves and solar wind flow.
Further complication comes from the finite wavelength effect,
     which introduces an apparent singularity at the Alfv\'en point where $M_T=1$ 
     for the wave equations~(\ref{eq_num_xi}) and (\ref{eq_num_eta}).
The numerical method employed for solving such an involved system of equations 
     is described in the next section.

\section{NUMERICAL IMPLEMENTATION}
\label{sec_num_model}

It is necessary to solve the governing equations described in section~\ref{sec_num_model}
     numerically for a quantitative analysis to be made.
To this end, the meridional magnetic field configuration
     should be prescribed,
     some external heat deposition needs to be applied to ions
     to generate the solar wind solutions,
     and appropriate boundary conditions need to be supplemented
     for both the wave and flow equations.
The implementation of these issues and a description of the method of solution
     are given in this section.

\subsection{Background Meridional Magnetic Field}
\label{sec_num_Bkgrnd_Mf}
For the meridional magnetic field, we adopt an analytical model given by 
     \citet{Bana_etal_98}.
{\revise In a spherical coordinate system, their Equation~(5b) may be rewritten as
\begin{eqnarray*}
&& \psi(r, \theta) 
=  M\left[\frac{\sin^2\theta}{r} 
    + \frac{3Q}{8}\frac{\sin^2\theta}{r^3}\left(4-5\sin^2\theta\right) \right .\nonumber \\
&&  +\left.\frac{K}{a_1}\left(1-\frac{a_1+r|\cos\theta|}{\sqrt{r^2+a_1^2+2 a_1 r|\cos\theta|}}\right)\right], \\
\end{eqnarray*}
where $\psi$ denotes the magnetic flux function, whose contours delineate the magnetic lines of force.
The poloidal magnetic field is then given by
\begin{eqnarray*}
\left[B_r, B_\theta\right] = \frac{1}{r\sin\theta}\left[
    \frac{1}{r}\frac{\partial \psi}{\partial\theta}, -\frac{\partial\psi}{\partial r}\right] .
\end{eqnarray*}}
In the present implementation, the model magnetic field consists of
     the dipole and current-sheet components only.
A set of parameters $M=2.9687$, $Q=0$, $K=0.9343$ and $a_1=1.5$ are
     chosen such that the last open magnetic field line is anchored
     at heliocentric colatitude $\theta=50^\circ$ on the Sun,
     while at the Earth orbit, the meridional magnetic field strength
     $B_l$ is 4$\gamma$ and independent of colatitude $\theta$,
     consistent with Ulysses measurements~\citep{SmithBalogh_95}.

\begin{figure}
\begin{center}
\epsscale{1.}
\plotone{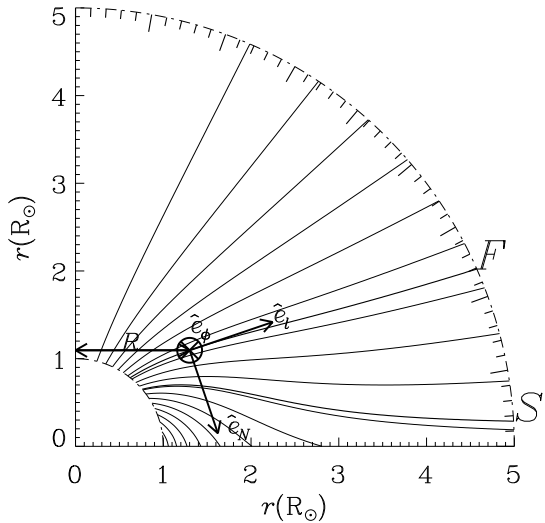}

\caption{
Adopted meridional magnetic field configuration in the inner corona.
Here only a quadrant is shown in which the magnetic axis points upward, and the thick countours 
     labeled $F$ and $S$ delineate the lines of force along which the fast
     and slow solar wind solutions are solved for, respectively.
Also shown is how to define the geometrical factor $R$,
     and the base vectors $\hat{e}_l, \hat{e}_N$ and $\hat{e}_\phi$ 
     of the flux tube coordinate system
     (see section~\ref{sec_model_descrip}).
}
\label{fig_mf}
\end{center}
\end{figure}

The background magnetic field configuration is depicted in Figure~\ref{fig_mf},
     where the thick contours labeled $F$ and $S$
     represent the lines of force along which we solve for
     fast and slow solar wind solutions, respectively.
They are also distinguished by the colatitudes $\theta_E$
     where they reach the Earth orbit $R_E=1$~AU in Table~\ref{tbl_1},
     which summarizes the parameters employed to generate the solar wind solutions.
Tube $F$ ($S$), which intersects the Earth orbit at 70$^\circ$ (89$^\circ$) colatitude,
     originates from $\theta=38.5^\circ$ ($49.4^\circ$)
     at the Sun where the meridional magnetic field strength $B_l$ is
     $5.24$ ($4.66$)~G.

\subsection{Ion Heating}
\label{sec_num_ion_heating}
The external energy deposition is assumed to come from
      an ad hoc energy flux that
      dissipates at a constant length $l_d$. 
The resulting total heating rate is therefore 
\begin{eqnarray*}
Q=F_{E} \frac{B_l}{B_{l E}l_d} \exp\left(-\frac{l}{l_d}\right),
\end{eqnarray*}
      where $F_E$ is the input flux scaled to
      the Earth orbit $R_E = 1$~AU, 
      $B_{l E}$ is the meridional magnetic field strength at $R_E$.
The heating rate $Q$ is then apportioned between protons and alpha particles
      according to 
\begin{eqnarray}
&& Q_p + Q_\alpha = Q, \hspace{0.2cm}
\frac{Q_\alpha}{Q_p} = \frac{\chi \rho_\alpha}{\rho_p}, \label{eq_num_adhoc_heating}
\end{eqnarray}
      where $\chi$ is a constant that indicates how alpha particles
      are favored when $Q$ is distributed, with
      $\chi=1$ standing for the neutral heating:
      the total energy that goes to ion species $k$ is proportional
      to its mass density $\rho_k$ ($k=p, \alpha$).
{\revise The choice of the heat deposition needs some explanation.
An exponential form of $Q$ was first suggested by \citet{HolzerAxford_70}
    and was later employed in a large number of studies.
The form of $Q$ adopted here is slightly different from the original version to ensure that
    it mimics the dissipation of a flux of non-thermal energy.
The way the dissipated energy is distributed
      resembles the mechanism involving ion-cyclotron waves
      (cf. \citet{IH83, HuHabbal_99}).}
Previous computations invoking such mechanisms indicate that,
      in the case of neutral heating, alpha particles tend to
      flow slower than protons (see the dispersionless case in Fig.1 of
      \citet{HuHabbal_99}).
Only when alpha particles are energetically favored in
      the corona can the modeled $U_{\alpha p}$ be positive
      in interplanetary space.
This happens when $\chi> 1$.
Listed in Table~\ref{tbl_1}, the heating parameters are chosen to yield
     fast and slow solar wind solutions with realistic ion mass fluxes and
     terminal speeds in the absence of wave contribution.
As a result, the influence introduced by the non-WKB Alfv\'en waves can be deduced.

\begin{table*}
\begin{center}
\caption{PARAMETERS USED TO GENERATE SOLAR WIND SOLUTIONS
\label{tbl_1}
}
\begin{tabular}{c|c|c c c|c c c}
\tableline
	&\multicolumn{1}{c|}{Magnetic Field} 
	&\multicolumn{3}{c|}{Heat Deposition}
	&\multicolumn{3}{c}{Base Flow Parameters} \\
\tableline
Wind	
	& $\theta_E$($^\circ$)		
	& $F_E$				& $l_d$			&$\chi$		 
	& $T_b$		& $n_{e, b}$			&$(n_\alpha/n_p)_b$	\\
Type
	& 
	& (erg cm$^{-2}$ s$^{-1}$)	& ($R_\odot$)		&  
	& ($10^6$~K)	& ($10^8$~cm$^{-3}$)	& \\
\tableline
Fast 	
	& 70 
	& 1.7 				& 1.6 			& 1.7 
	& 1		& 3			&6\%			\\
slow 	
	& 89
	& 1.2 				& 1.4 			& 4 
	& 1.4		& 3			& 6\%	\\		
\tableline
\end{tabular}
\end{center}
\end{table*}

\subsection{Boundary Conditions and Method of Solution}
\label{eq_num_method_of_solution}
The governing equations consisting of the flow part (Eqs.(8) to (10) in paper I) and
      the wave part (Eq.(38) in paper I) are solved on a spatial grid extending from
      the coronal base (1~$R_\odot$) out to 100~$R_\odot$.
Both parts need to be supplemented with appropriate
     boundary conditions.
For the flow part, at the base 1~$R_\odot$ the flow speeds
     $U_{p}$ and $U_{\alpha}$ are determined by mass conservation,
     whereas the ion densities as well as species temperatures are fixed.
These parameters are given in Table~\ref{tbl_1}, where the ion densities are given in terms of 
     the base values of the electron density $n_{e,b}$ and the alpha particle
     abundance $(n_\alpha/n_p)_b$.
At the top boundary (100~$R_\odot$), all the flow parameters are linearly extrapolated for simplicity.
On the other hand, for a monochromatic wave with angular frequency $\omega$,
     the regularity requirement at the Alfv\'en point 
     means that the wave acceleration $a_{k}$ is determined only 
     by one parameter, which
     is chosen to be the electron velocity fluctuation amplitude $\left<u_e^2\right>^{1/2}$ at 1~$R_\odot$,
     or $\delta u_b$ for brevity.
To evaluate the non-WKB effect, we also construct models with WKB waves with the same base flow conditions
     and identical base wave amplitudes.

We adopt an iterative approach to solve simultaneously the nonlinear system of equations.
With an initial guess for the flow parameters, the wave acceleration $a_{k}$ ($k=p, \alpha$)
     may be obtained as follows.
When a non-WKB Alfv\'en wave with a given $\omega$ is considered,
     equations~(\ref{eq_num_xi}) and (\ref{eq_num_eta}) are first solved analytically at the Alfv\'en point,
     and then integrated both inward to the coronal base and outward to 100~$R_\odot$
     using the standard 4-th order Runge-Kutta method. 
One may find $a_{k}$ through Equation~(\ref{eq_def_wave_acce}).
When constructing models with WKB waves, the amplitude of the magnetic field fluctuation $|b|$ is first found 
     at the base from relation~(\ref{eq_wkb_eigen}), and then computed via
     Equation~(\ref{eq_wkb_actioncons}) for the whole computational domain.
The wave acceleration $a_{k}$ is readily obtained from Equation~(\ref{eq_wkb_acce}).
The flow equations (Eqs.(8) to (10) in paper I) incorporating $a_{k}$ are then solved
    by using the numerical scheme devised by \citet{Hu_etal_97},
    thereby all the flow parameters are updated.
To ensure internal consistency, the two steps are iterated until a convergence is met.
This iterative approach closely follows those used by
    \citet{AlazrakiCouturier_71} and \citet{MacChar_94},
    where WKB and non-WKB Alfv\'en waves are self-consistently 
    included in single-fluid solar wind models, respectively.

\section{NUMERICAL RESULTS}
\label{sec_num_res}

Having described the solution method, we
     are now in a position to answer the following questions: 
     to what extent the wind parameters are affected by the non-WKB effect in general?
     and how the Helium abundance and the proton--alpha particle differential speed are 
     affected in particular?
To address these questions, we will only vary those parameters characterizing the waves,
     namely, the base wave amplitude $\delta u_b$,
     and angular frequency $\omega$.
{\revise There has been considerable effort made to determine $\xi$, the non-thermal motions
     that contribute to the broadening of some UV lines,
     in the solar upper atmosphere by using SUMER and UVCS data
     \citep[e.g.,][]{Banerjee_etal_98, Chae_etal_98, Peter_99a, Esser_etal_99}.
Typically $\xi$ is found to be $\sim 30$\velunits\ 
     at an altitude of $\sim 0.02$~$R_\odot$ above limb.
Since this value corresponds to the line-of-sight component, the overall magnitude
     of the non-thermal motion may be $\sqrt{2}\xi$ for circularly polarized waves
     \citep[cf.][]{Banerjee_etal_98, CranmerBalle_05}.
Given that the waves examined here are linearly polarized, $\delta u_b$ will be allowed to have 
     a value of up to $40$\velunits.}
In this section, we will fist examine the fast and then
     the slow solar wind.

\subsection{Fast Solar Wind Solutions}
\begin{figure}
\begin{center}
\epsscale{1.}
\plotone{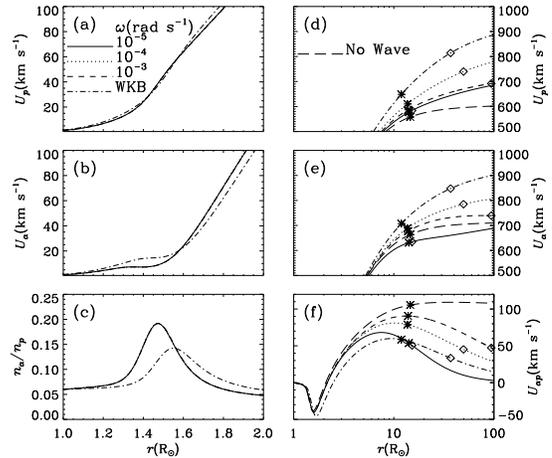}
\caption{
Multicomponent solar wind models incorporating non-WKB Alfv\'en waves.
Numerical results for three angular frequencies and for the WKB wind model
     are given in different line styles as indicated in (a).
All models have the same base velocity fluctuation amplitude $\delta u_b=28$\velunits.
Radial profiles are shown for 
     (a) and (d) the proton flow speed $U_p$,
     (b) and (e) the alpha flow speed $U_\alpha$,
     (c) the alpha abundance $n_\alpha/n_p$,
     and 
     (f) The proton-alpha differential speed $U_{\alpha p}=U_\alpha -U_p$.
Panels in the left column  give a close-up of the inner corona $r\le 2$~$R_\odot$.
In (d), (e) and (f), the long-dashed curves correspond to the waveless model,
     the asterisks indicate the location of the Alfv\'enic point, defined by Equation~(\ref{eq_def_MT})
     to be the location where the combined
     Alfv\'enic Mach number $M_T=1$.
The diamonds mark the location where the wave-induced magnetic field fluctuation amplitude
     starts to exceed the background.
}
\label{fig_fast_flow}
\end{center}
\end{figure}

Figure~\ref{fig_fast_flow} examines the flow properties of the fast solar wind solutions
     corresponding to several $\omega$ as plotted by different line styles indicated
     in Figure~\ref{fig_fast_flow}a.
For comparison, also displayed are the solution incorporating WKB waves (the dash-dotted curves)
     and that without wave contribution (long-dashed curves, only shown in the right column).
The models involving waves all pertain to a wave amplitude of
      $\delta u_b = 28$~\velunits.
The left column gives the radial distributions near the base of the flow $r\le 2$~$R_\odot$ of 
      the flow speeds of (a) protons $U_p$ and (b) alpha particles $U_\alpha$,
      as well as that of (c) the alpha particle abundance
      $n_{\alpha p} = n_\alpha/n_p$. 
In addition, the right column gives the radial profiles
      for the entire computational domain of 
      (d) $U_p$, (e) $U_\alpha$,
      and (g) the difference between the two $U_{\alpha p}=U_\alpha-U_p$.
The asterisks in the right column mark the Alfv\'en point, where
      the combined meridional Alfv\'enic Mach number $M_T=1$, with
      $M_T$ defined by Equation~(\ref{eq_def_MT}).
Moreover, for the solutions with waves
      the diamonds correspond to the location $r_{NL}$ where 
      the amplitude of the wave-induced magnetic field fluctuations
      equals the background, i.e., $\left<b^2\right>/B_l^2 = 1$.
Care has to be taken when one examines the segments
      at distances beyond those diamonds
      since when deriving the wave equation,
      we assume that the time-dependent solar wind flow may be decomposed into a steady background
      and Alfv\'en waves.
One may expect that when the wave amplitudes are significant such a decomposition is not permitted.
However, as first discussed by \citet{Lou_93}, the source terms introduced into the momentum
      and magnetic induction equations by
      wave-induced fluctuations decrease sufficiently fast
      with distance asymptotically.
Consequently, the first-order wave amplitudes are valid provided that the wave amplitude imposed at the coronal base
      is sufficiently small. 
Nevertheless, the portions where $\left<b^2\right>/B_l^2 \ge 1$ are indicated for clarity, and 
      we have excluded all solutions for which $r_{NL} < r_A$ so that
      integrating the wave equation backwards from $r_A$ to the coronal base is formally allowed.

\begin{figure}
\begin{center}
\epsscale{1.}
\plotone{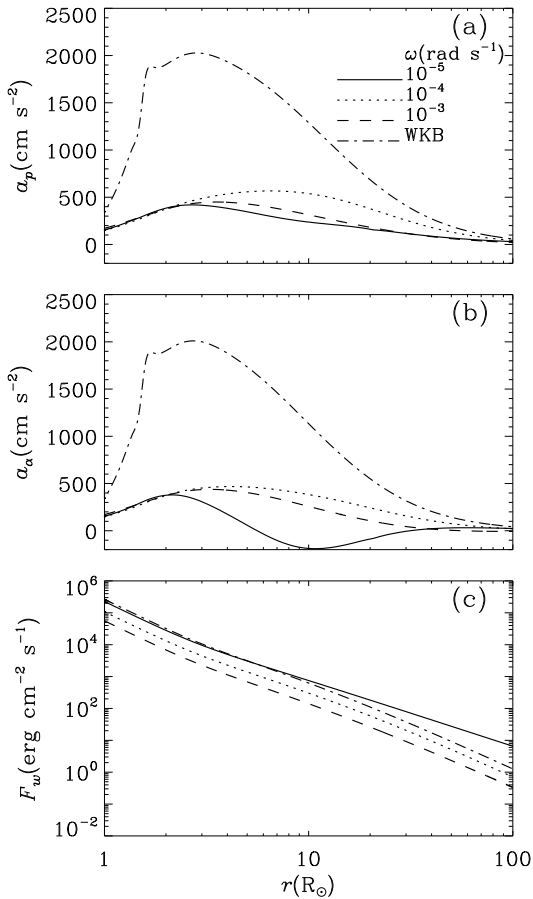}
\caption{
Properties of non-WKB Alfv\'en waves in self-consistently constructed multicomponent solar wind models.
Shown are the radial distributions of the wave acceleration experienced by (a) protons and (b) alpha particles,
     as well as (c) the wave energy flux density
     for three angular frequencies and the WKB wind model
     as given in different line styles indicated in (a).
All models have a common base velocity fluctuation amplitude $\delta u_b=28$\velunits.
}
\label{fig_fast_wave}
\end{center}
\end{figure}

A close examination of the left column of Figure~\ref{fig_fast_flow} reveals that
      the non-WKB models are virtually indistinguishable.
Actually they differ only slightly from the waveless model, indicating that
      the wave acceleration associated with these frequencies plays an insignificant role
      in the force balance for protons or alpha particles in the region considered.
On the other hand, an obvious difference exists  between the WKB and non-WKB wind models.
For instance, Figures~\ref{fig_fast_flow}a and \ref{fig_fast_flow}b demonstrate that
      the WKB one tends to produce a higher 
      flow speed for both protons and alpha particles near the base,
      and hence a higher ion flux as the ion densities are fixed at 1~$R_\odot$.
Consider the proton and alpha fluxes scaled to 1~AU, $(n_p U_p)_E$ and $(n_\alpha U_\alpha)_E$.
{In the WKB case $(n_p U_p)_E$ ($(n_\alpha U_\alpha)_E$) is 
      $3.18\times 10^{8}$ ($1.49\times 10^{7}$)\nofluxunits, 
      whereas in the model with $\omega=10^{-5}$\angfrequnits, $(n_p U_p)_E$ and $(n_\alpha U_\alpha)_E$
      are $2.6\times 10^{8}$ and $1.14\times 10^{7}$\nofluxunits, respectively.}
Furthermore, as shown in Figure~\ref{fig_fast_flow}c, the WKB case corresponds to a reduced
      alpha abundance $n_{\alpha p}$ for $r\lesssim 1.56$~$R_\odot$,
      with the local maximum $n_{\alpha p, M}$ decreasing from $0.19$ attained at
      $1.47$~$R_\odot$ in the non-WKB models to $0.14$ at $1.55$~$R_\odot$ in the WKB one.
This change may be readily understood since compared with their non-WKB counterparts, the WKB waves
      produce an enhancement of the alpha speed $U_\alpha$ larger than that of the proton one $U_p$.
As a result, at the base of the wind the WKB model yields a smaller $U_p/U_\alpha$ and hence a smaller
      $n_{\alpha}/n_p$ since the ion flux ratio $(n_\alpha U_\alpha)/(n_p U_p)$ is a constant
      for steady flows.
An interesting aspect of Figure~\ref{fig_fast_flow}b is 
      that the alpha flow speed $U_\alpha$ in the WKB model
      becomes smaller than in the non-WKB ones for $r\gtrsim 1.58$~$R_\odot$,
      despite that the WKB wave acceleration exerted on alpha particles
      $a_{w, \alpha}$ is substantially larger
      than the non-WKB ones throughout the region considered (see Fig.\ref{fig_fast_wave}b). 
This is because close to the coronal base, the reduced alpha abundance $n_{\alpha}/n_p$
      in the WKB model gives rise to a lower heating rate for alpha particles (cf. Eq.(\ref{eq_num_adhoc_heating}))
      and therefore a lower alpha temperature.
Hence the lowered alpha pressure gradient force results in a $U_\alpha$ profile that is less steep
      at $r\gtrsim 1.42$~$R_\odot$.

Now turn to the right column, from which one may gain a first impression of the overall
       influence introduced by the non-WKB effect on the flow parameters.
Take the variation of the location of the Alfv\'en point, $r_A$, for example.
For $\omega=10^{-5}, 10^{-4}$ and $10^{-3}$\angfrequnits, 
       $r_A$ is $14$, $13.6$, and $13.9$~$R_\odot$, respectively.
These values are close to $r_A=14.6$~$R_\odot$ derived for the waveless model.
On the other hand, $r_A$ is $11.9$~$R_\odot$ in the WKB model, reinforcing the fact
      that the WKB Alfv\'en waves are more effective than the non-WKB ones in accelerating the solar wind flow
      in the sub-Alfv\'enic portion.
However, Figure~\ref{fig_fast_flow}d indicates that even in the WKB case, 
      the proton speed $U_{p}$ is only $648$\velunits\ at $r_A$, which is only slightly enhanced 
      compared with the model without waves, for which $U_p=577$\velunits\ at the Alfv\'en point.
More significant difference arises in the super-Alfv\'enic portion, where protons receive 
      a continuous acceleration from both the WKB and non-WKB Alfv\'en waves.
Furthermore, the proton flow speed does not show a monotonic dependence
      on wave frequency.
For instance, at 100~$R_\odot$, the WKB wind model yields a $U_p=887$\velunits\, whereas
      the non-WKB models yield $U_{p}=686, 778$ and $694$\velunits\ for
      $\omega=10^{-5}, 10^{-4}$ and $10^{-3}$\angfrequnits\, respectively.
For comparison, the waveless model obtains $602$\velunits\ for $U_p$ at the same location.

Inspection of Figure~\ref{fig_fast_flow}e shows that among the non-WKB models considered, 
      the one with $\omega=10^{-5}$\angfrequnits\ shows a fundamental difference from the
      rest as far as the alpha flow speed $U_\alpha$ is concerned:
      this wave tends to decelerate rather than accelerate the 
      alpha particles.
In fact, for the entire computational domain $U_\alpha$ is even smaller than that in the waveless model.
In contrast, similar to the WKB case, the non-WKB models with $\omega=10^{-4}$ and $10^{-3}$\angfrequnits\
      produce a higher $U_\alpha$ compared with the model without waves.
The disparate effects on the ion flows of fluctuations with $\omega=10^{-4}$
      and $10^{-5}$\angfrequnits\ signify a transition
      around some $\omega_c$ below which the waves behave in a quasi-static manner,
      a feature extensively discussed by~\citet{HO80} and \citet{Lou_93}
      in the case of single-fluid MHD.
Given a background flow, $\omega_c$ in the present case may be approximated by
\begin{eqnarray}
 \omega_c \approx U_{mA}/(2 r_A) ,
\label{eq_def_omegac}
\end{eqnarray}
      where $U_{mA}$ is the speed of the center of mass evaluated at the Alfv\'en point.
Note that when the non-WKB waves are self-consistently incorporated, the RHS of Equation~(\ref{eq_def_omegac})
      may depend on the wave frequency and base amplitude.
However, it turns out that the dependence is rather weak, a value of $\omega_c \sim 3\times 10^{-5}$\angfrequnits\
      applies for all the fast solar wind solutions considered.
For those $\omega>\omega_c$, one may expect that the waves will be increasingly WKB-like
      with increasing $\omega$, whereas for $\omega <\omega_c$, the effects of quasi-static fluctuations
      are similar to those of solar rotation, the zero-frequency solution to the wave equation.

Whether or not a quasi-static behavior results, the net effect of Alfv\'en waves
      on the speed difference $U_{\alpha p}$
      is to reduce its magnitude, as evidenced by Figure~\ref{fig_fast_flow}f.
{For instance, for $\omega=10^{-5}$\angfrequnits one finds that 
      $U_{\alpha p}=$ is 2.87\velunits at 100~$R_\odot$, where
      the corresponding values for $\omega=10^{-4}$ and $10^{-3}$\angfrequnits\ are
      $28.2$ and $45.5$\velunits, respectively.
As for the WKB model, one finds that $U_{\alpha p}= 14.7$\velunits.
All these values are considerably smaller than $108$\velunits, found at the same location in the waveless model.}
This is not surprising since it is readily shown that if the flow speeds are
      entirely determined by the wave forces, then 
\begin{eqnarray}
 \left(U_{\alpha}^2-U_p^2\right)\left[1 + \epsilon \frac{\left<b^2\right>}{B_l^2}\right] = \mbox{const},
\label{eq_wave_speeddiff}
\end{eqnarray}
      where the factor $\epsilon =1$ for WKB waves.
On the other hand, relation~(\ref{eq_wave_speeddiff})
      also holds when $\omega$ approaches zero
      (cf. \citet{LiLi06}), the only difference is that $\epsilon$ should be chosen to be $2$.
For a wave with an arbitrary finite $\omega$, it is natural to expect that the effect lies between the two extremes.
Now that the ratio $\left<b^2\right>/B_l^2$ increases monotonically with increasing $r$
      for all the wave models considered,
      all models should yield a $U_{\alpha p}$ decreasing with $r$ if $U_\alpha+U_p$
      does not show a substantial variation,
      as confirmed by the super-Alfv\'enic portions of the flow.

The behavior of the flow parameters is further explained by
      Figure~\ref{fig_fast_wave}, which displays
      the radial profiles of the wave acceleration inflicted
      on (a) protons $a_{p}$ and (b) alpha particles $a_{\alpha}$, as well as
      (c) the wave energy flux density $F_w$.
Numerical results are shown for the three angular frequencies considered, and the WKB model.
Figures~\ref{fig_fast_wave}a and \ref{fig_fast_wave}b show that within $\sim 2$~$R_\odot$
      the non-WKB wave acceleration differs little from one another,
      but is considerably smaller than that in the WKB case.
Specifically, one finds that at 2~$R_\odot$, for $\omega=10^{-5}, 10^{-4}$ and $10^{-3}$\angfrequnits, $a_{p}$ is
       $384$, $375$ and $373$\accunits, respectively, while the WKB model
       yields a value of $1904$\accunits.
At larger distances, the difference between the WKB and non-WKB models is also prominent.
For instance, at 100~$R_\odot$, one finds that $a_{p}=58.9$\accunits\ for the WKB model, while
       $30.2, 42.7,$ and $21.7$\accunits\ for $\omega=10^{-5}, 10^{-4}$ and $10^{-3}$\angfrequnits, respectively.
The nonmonotonic frequency dependence of $a_{p}$ is the direct reason for the similar dependence of
       the proton speed $U_p$.
Likewise, the reduction of the alpha speed $U_\alpha$ for $\omega=10^{-5}$\angfrequnits\
       relative to the waveless model results from the fact that in a large portion
       of the computational domain, $5.7\lesssim r/R_\odot \lesssim 31.1$ to be precise,
       the wave exerts a negative $a_{\alpha}$ on alpha particles.
For $\omega=10^{-3}$ and $10^{-4}$\angfrequnits, however, the waves tend to 
       accelerate the alpha particles, as in the WKB case.
Note that in relation~(\ref{eq_wave_speeddiff}), the wave effect in limiting
      $|U_{\alpha p}|$ does not rely on the sign of $U_{\alpha p}$.
Hence we may conclude that the reduction of $|U_{\alpha p}|$
      is achieved in different manners by wave-like and quasi-static fluctuations.
While wave-like fluctuations tend to provide overall acceleration for both ion species,
      the quasi-static ones may accelerate the slower flowing species but decelerate
      the species that flows faster.
{It is noted, however, that the conclusion is based on the condition that $|U_{\alpha p}| \lesssim U_A$, which
      holds for both the fast and slow solar wind as measured by Helios~\citep{Marsch_etal_82a},
      and is also true for all the wave-based solutions examined here.
For $|U_{\alpha p}| \gtrsim U_A$ kinetic instabilities may arise when the proton beta is comparable to unity
      and the physics is more complicated
      \citep[e.g.][]{Gary_etal_00}.
If $|U_{\alpha p}|$ is larger still, then from Equation~(\ref{eq_wkb_disprel}) one may deduce that 
      a low frequency electromagnetic Alfv\'en instability will set in
      \citep[cf.][]{Verheest_77}.
Both subjects are beyond the scope of the present paper.}

To proceed, we note that the governing equations allow an energy conservation
       law to be derived, from which it is readily recognized that
       at large distances $r\gg r_A$,
       the gain in the ion kinetic and gravitational potential energy fluxes
       derives primarily from the energy fluxes associated with waves and the adopted ion heating.
In other words, by noting that at $r\gg r_A$ the empirical energy flux has been exhausted, 
       one may deduce the overall contribution of Alfv\'en waves between 1~$R_\odot$ and $r$ from
       the relation
\begin{eqnarray}
&& \sum_{k=p, \alpha} \left(\frac{\rho_k U_k B_{lE}}{B_l}\right)
        \frac{U_k^2 (r) + V_{esc}^2}{2}\nonumber \\
&& \approx W(R_\odot) - W(r) + F_{E}, 
\label{eq_enerbalance}
\end{eqnarray}
       where $W = (B_{lE}/{B_l}) F_w$ denotes the wave energy flux scaled to the Earth orbit.
Moreover, $V_{esc}=\sqrt{2G M_\odot/R_\odot}$ is the escape speed.
Equation~(\ref{eq_enerbalance}) indicates that if the ion fluxes in different models are similar, then 
       the asymptotic ion speed $U_k$ is largely determined by the difference of $W$ between
       1~$R_\odot$ and $r$, which hereafter will be taken as 100~$R_\odot$.
Moreover, at large distances $U_\alpha$ is close to $U_p$ in models with waves.
Therefore, a frequency dependence of $\Delta W$ similar to that of $U_p$ is expected and indeed reproduced
       in Figure~\ref{fig_fast_wave}c.
For instance, $\Delta W = W(R_\odot) - W(100 R_\odot)$ is found to be $0.31$, $0.67$ and $0.36$ (here and
       hereafter in units of \funits) for
       $\omega=10^{-5}, 10^{-4}$ and $10^{-3}$\angfrequnits, respectively.
On the other hand, $\Delta W=1.75$ is considerably larger in the WKB model.
In fact, this value more than compensates for the enhancement of the ion speeds
       in that among the examined ion fluxes
       the WKB model also {\revise yields the largest value}.
An interesting feature is that unlike the frequency dependence of $\Delta W$, 
       at 1~$R_\odot$ the wave energy flux density for $\omega=10^{-5}$\angfrequnits\
       is the closest to the corresponding WKB value.
Specifically, when scaled to $R_E$, the WKB model yields that $W(R_\odot)=2.02$, whereas
        the values for $W(R_\odot)$ are $1.73$, $0.82$ and
        $0.43$ for $\omega=10^{-5}, 10^{-4}$ 
        and $10^{-3}$\angfrequnits, respectively.
From the definition of the wave energy flux density (cf. Eq.(20b) in paper I), it is easily understood
        that this feature comes largely from the frequency dependence
        of the wave-induced magnetic fluctuation $b$.
For $\omega=10^{-5}, 10^{-4}, 10^{-3}$ and the WKB model,
        $|b|$ is found to be $0.28$, $0.15$, $0.082$ and $0.32$~G, respectively.
Now examine the fraction $\Delta W/W(R_\odot)$. 
One may see that
        only 18.1\% of the injected wave flux is lost in the form of the work done on the ion flows
        for $\omega=10^{-5}$\angfrequnits,
        whereas for the WKB model the corresponding fraction is 86.4\%.
As for the model with $\omega=10^{-4}$ ($10^{-3}$)\angfrequnits, this fraction
       is 81.4\% (83.2\%), close to that found in the WKB case.
Therefore one may envision that $\omega_c$ also distinguishes fluctuations in their efficiency
       of losing energy to the solar wind flow, with the quasi-static ones being much less efficient.

One may have noticed that in the cases with $\omega=10^{-3}$ and $\omega=10^{-4}$ as well as in the WKB model
        the $F_w$ profiles are rather similar. 
As has been discussed in paper I, in the near-Sun regions the $F_w$ profiles
        for all the examined waves roughly behave like $F_w \propto B_l$, indicating that the wave energy
        is diluted only by the expansion of the magnetic flux tube.
On the other hand, at large distances $r\gg r_A$, the $F_w$ profiles for $\omega=10^{-3}$ and $10^{-4}$\angfrequnits\
        roughly comply with the WKB expectation, i.e., $F_w \sim r^{-3}$, since in the region considered
        the WKB limit applies in these two cases.
When it comes to the model where $\omega=10^{-5}$\angfrequnits,
        one may find that asymptotically $F_w \sim r^{-2}$ approximately.

\begin{figure*}
\begin{center}
\epsscale{1.}
\plotone{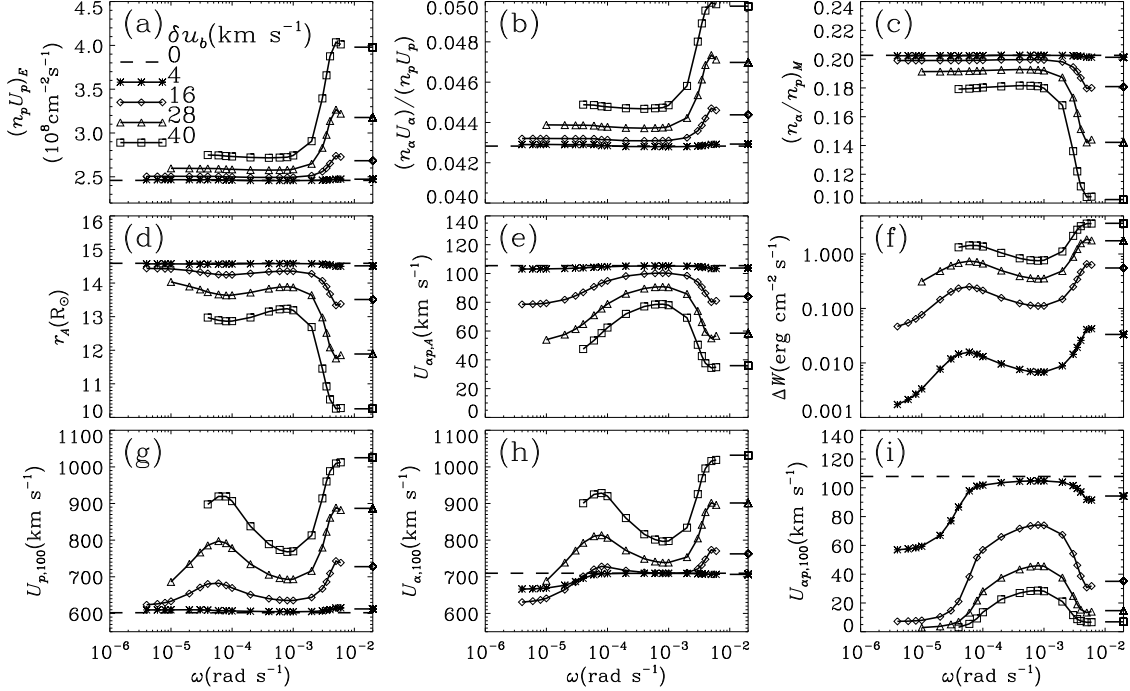}
\caption{
Frequency dependence of several parameters characterizing the multicomponent solar wind solutions
       self-consistently incorporating non-WKB Alfv\'en waves.
The four series of models pertain to different base velocity
       fluctuation amplitudes as indicated in (a).
In addition, the dashed lines give the waveless model, i.e., $\delta u_b=0$.
(a) The proton flux scaled to the Earth orbit $(n_p U_p)_E$.
(b) The ratio of the alpha flux to the proton one $(n_\alpha U_\alpha)/(n_p U_p)$.
(c) The local maximum in the inner corona of the alpha abundance $(n_\alpha/n_p)_M$.
(d) The location of the Alfv\'en point $r_A$.
(e) The proton-alpha speed difference at the Alfv\'en point, $U_{\alpha p, A}$.
(f) The difference of $W$, the wave energy flux density scaled to the Earth orbit, between 1 and 100~$R_\odot$.
(g), (h), and (i) The proton $U_{p, 100}$ and alpha $U_{\alpha, 100}$ flow speeds as well
    the proton-alpha speed difference $U_{\alpha p,100}$ at 100~$R_\odot$.
For comparison, the parameters derived in the corresponding WKB wind models are
    given by the short horizontal bars in each panel.
}
\label{fig_fast_par}
\end{center}
\end{figure*}

Figure~\ref{fig_fast_par} expands the obtained results by 
        displaying the dependence on the angular frequency $\omega$
        of some flow parameters for waves with four different base
        amplitudes $\delta u_b = 4, 16, 28$ and $40$\velunits.
For comparison, the waveless model is shown by the dashed lines,
        while the corresponding WKB results are given by the horizontal bars at the right of each panel.
Plotted in Figure~\ref{fig_fast_par} are (a) the proton flux scaled to the Earth orbit $(n_p U_p)_E$,
        (b) the ion flux ratio $(n_\alpha U_\alpha)/(n_p U_p)$,
        (c) the local maximum of the alpha abundance represented by $n_{\alpha p, M}$,
        (d) the location of the Alfv\'en point $r_A$,
        (e) the proton-alpha differential speed $U_{\alpha p, A}$ at the Alfv\'en point,
     and (f) the loss of the wave energy between 1~$R_\odot$ and 100~$R_\odot$, denoted by $\Delta W$.
Also given are the values at 100~$R_\odot$ for
        (g), (h) and (i) the proton and alpha speeds, $U_{p,100}$ and $U_{\alpha, 100}$, as well as
              the proton-alpha speed difference $U_{\alpha p, 100}$.
Let us consider Figures~\ref{fig_fast_par}a to \ref{fig_fast_par}e first.
One can see that as far as the the wave effects inside the Alfv\'en point are concerned,
        for $\omega \gtrsim 3.5\times 10^{-3}$\angfrequnits,
        the relative difference of the computed parameters with respect to the WKB values
        is less than 10\%.
In other words, the WKB limit provides an adequate description for these high frequency waves.
However, for lower frequencies significant differences
         tend to appear between the non-WKB and WKB models, with the latter being always more effective
         in accelerating the solar wind as indicated by the smaller $r_A$
         (see Fig.\ref{fig_fast_par}d). 
On the other hand, Figure~\ref{fig_fast_par}b indicates that the difference between the WKB and non-WKB
         values for the ion flux ratio is rather modest, 
         even for $\delta u_b=40$\velunits\ the relative difference
         is $\lesssim 10.2\%$.
Given that the proton flux is substantially enhanced in the WKB case (see Fig.\ref{fig_fast_par}a),
         this modest difference in the ion flux ratio means that relative to the non-WKB waves,
         the WKB ones boost the fluxes of both ion species to nearly equal extent.
Note that this does not guarantee that the alpha abundance in the inner corona is 
         only slightly modified by the non-WKB effect.
In fact, the local maximum $n_{\alpha p, M}$, which 
         varies from $0.168$ to $0.182$ for $4\times 10^{-5} \le \omega \le 2\times 10^{-3}$\angfrequnits\
         in the series where $\delta u_b = 40$\velunits,
         is up to $77.3\%$ larger than $n_{\alpha p, M}=0.102$ obtained in the WKB model.
As for the proton-alpha speed difference, panel (e) shows that the waves with $\delta u_b \gtrsim 16$\velunits\
         have appreciable effect in limiting its magnitude already inside the Alfv\'en point.
Moreover, the wave with $\omega=8\times 10^{-4}$\angfrequnits\ seems to be the most inefficient
          in achieving this effect in each series.
Now consider Figures~\ref{fig_fast_par}f to \ref{fig_fast_par}i, 
         where the wave effects on the asymptotic flow parameters are examined.
It can seen in Figure~\ref{fig_fast_par}f that in general the asymptotic proton speed $U_p$ is correlated with the
        wave energy loss $\Delta W$, a consequence of relation~(\ref{eq_enerbalance}).
However, the correlation disappears for the lowest wave amplitude in the segment
         $\omega \lesssim 5\times 10^{-5}$\angfrequnits.
This is not surprising since in this case the wave energy loss $\Delta W$ is too low to play a role.
As a matter of fact, in this frequency range $\Delta W$ is $\lesssim 0.93\%$ of $F_E$, the
         energy flux associated with the ion heating.
However, despite this, the waves show appreciable effects in limiting the proton-alpha
        speed difference (see Fig.\ref{fig_fast_par}i).
Specifically, in the waveless model, $U_{\alpha p, 100}=108$\velunits,
        whereas in the model with $\omega=10^{-5}$\angfrequnits,
        $U_{\alpha p, 100}$ is $59.4$\velunits.
The reduction is seen to be achieved by a reduction in $U_\alpha$ and an increase in $U_p$
        (Figs.\ref{fig_fast_par}g and \ref{fig_fast_par}h).
Comparing Figure~\ref{fig_fast_par}i with \ref{fig_fast_par}e, one can see that between $r_A$ and 100~$R_\odot$,
        $U_{\alpha p}$ is further reduced.
Furthermore, the feature that the waves with $\omega=8\times 10^{-4}$\angfrequnits\ correspond to the largest
        $U_{\alpha p}$ for a given series also shows up in Figure~\ref{fig_fast_par}i.
At the higher and lower ends of the frequency range, $U_{\alpha p, 100}$ nearly vanishes
        for wave amplitudes $\delta u_b \gtrsim 16$\velunits.

\subsection{Slow Solar Wind Solutions}

Let us now move on to discuss the slow solar wind, or rather,
       how Alfv\'en waves may influence the flow parameters in models
       obtained along tube $S$ indicated in Figure~\ref{fig_mf}
       using the heating and boundary flow parameters corresponding to ``slow wind''
       in Table~\ref{tbl_1}.
For the chosen parameters, the waveless model yields a proton flux of $(n_p U_p)_E$ of
       $3.73 \times 10^{8}$\nofluxunits\,
       and an ion flux ratio of $(n_\alpha U_\alpha)/(n_p U_p)=0.04$.
A local maximum of the alpha abundance $n_{\alpha p, M}=0.185$ is found at 1.9~$R_\odot$.
The flow reaches the Alfv\'en point at 17.2~$R_\odot$, where the proton (alpha) speed is
       258 (367)\velunits, the resulting speed difference being 109\velunits.
At 100~$R_\odot$, $U_{p, 100}= 290$\velunits and $U_\alpha=370$\velunits.
It is noteworthy that the corresponding $U_{\alpha p}=80$\velunits\ is 
       not unrealistic for slow solar winds, even larger values have been found
       by Helios 2 when approaching perihelion~\citep{Marsch_etal_81}.

\begin{figure*}
\begin{center}
\epsscale{1.}
\plotone{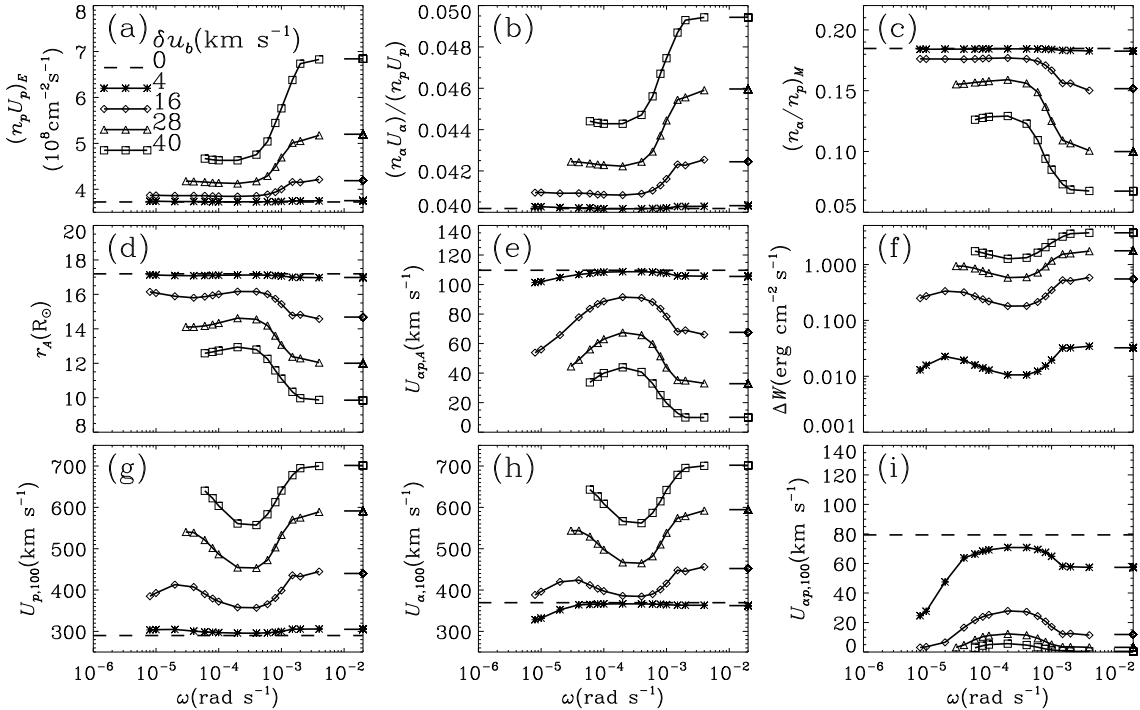}
\caption{
Similar to Figure~\ref{fig_fast_par} but for the slow solar wind solutions.
}
\label{fig_slow_par}
\end{center}
\end{figure*}

Figure~\ref{fig_slow_par} presents, in the same format as Figure~\ref{fig_fast_par}, the frequency dependence
       of several flow parameters obtained in four series of solar wind models with different
       $\delta u_b$.
In addition, the WKB and waveless models are also given.
A comparison with Figure~\ref{fig_fast_par} indicates that nearly all the features in
       Figure~\ref{fig_slow_par} are reminiscent of those obtained for
       fast solar wind solutions.
In particular, Figures~\ref{fig_slow_par}a and \ref{fig_slow_par}d indicate that the non-WKB waves produce
       a proton flux up to $32.3\%$ lower, a value for the location of the Alfv\'en point up to $31.2\%$ higher
       than their WKB counterparts.
As in the case of fast wind solutions, the WKB and non-WKB Alfv\'en waves show only a modest difference
       in the derived ion flux ratio, indicating that the larger wave accelerations in the WKB case
       boost the fluxes of both species to similar degrees
       (see Fig.\ref{fig_slow_par}b).
However, for the local maximum alpha abundance $n_{\alpha p, M}$ in the inner corona the non-WKB values 
       may be up to $91.5\%$ larger (Fig.\ref{fig_slow_par}c).
Furthermore, Figure~\ref{fig_slow_par}e indicates that for 
       $\delta u_b\gtrsim 16$\velunits\ both WKB and non-WKB waves may substantially reduce $U_{\alpha p}$
       inside $r_A$.
Note that the waves with $\omega\sim 2\times 10^{-4}$\angfrequnits\ are the most inefficient in achieving
       this effect, and the same feature persists in Figure~\ref{fig_slow_par}i, which
       shows that $U_{\alpha p}$ at 100~$R_\odot$ is no larger than 27.9\velunits\ for all
       models with $\delta u_b\gtrsim 16$\velunits.
Interestingly, at this distance, the waves with the lowest frequency may also be effective even with
       the lowest amplitude.
In fact, for $\omega = 8\times 10^{-6}$\angfrequnits, the waves with $\delta u_b=4$\velunits\ yield that
       $U_{\alpha p, 100}=24.6$\velunits, even though the corresponding wave energy loss is
       merely $0.013$\funits, amounting to $1.1\%$ of the empirical energy flux exhausted to heat ion fluids
       (cf. Fig.\ref{fig_slow_par}f).
In this particular case, the wave energy flux plays no part in the solar wind acceleration, 
       rather waves provide a net acceleration for the protons but a net deceleration for the alphas
       (cf. Figs.\ref{fig_slow_par}g and \ref{fig_slow_par}h).
Moreover, Figures~\ref{fig_slow_par}g and \ref{fig_slow_par}h indicate that for $\delta u_b = 16$\velunits,
       compared with the waveless values $U_\alpha$ for $\omega = 8\times 10^{-6}$\angfrequnits\ is barely larger,
       but $U_p$ shows a significant increase by $95$\velunits.
On the other hand, for the WKB and the majority of the non-WKB waves, both $U_p$ and $U_\alpha$
       are enhanced relative to the waveless values.
This once again reinforces the notion that different mechanisms operate at different frequency regimes.
A further comparison with Figures~\ref{fig_fast_par} indicates that 
       some quantitative differences arise in Figure~\ref{fig_slow_par} nonetheless.
One may see that the non-WKB values approach the corresponding WKB one at a lower frequency.
To be more specific, the WKB limit provides a description with relative differences
       within $10\%$ for the wave behavior as long as $\omega \gtrsim 1.5\times 10^{-3}$\angfrequnits.
This happens in conjunction with the lowering of the critical value $\omega_c$ below which the waves behave
       in a quasi-static rather than a wave-like way.
For the solutions examined, a value of $1.3\times 10^{-5}$\angfrequnits\ may be quoted for $\omega_c$.

\section{SUMMARY AND CONCLUDING REMARKS}
\label{sec_conc}

This study has been motivated by the observational facts that Alfv\'enic fluctuations exist in both
        fast and slow solar winds~\citep{Marsch_etal_81, Marsch_etal_82a, Marsch_etal_82b}, and 
        for the majority of the measured fluctuations the corresponding frequency is too low to 
        allow the short wavelength WKB limit to apply~\citep[e.g.,][]{TuMarsch_95}. 
Specifically, using the formulation given in \citet{LiLi07},
        we have constructed multicomponent solar wind models
        that treat protons and alpha particles on an equal footing
        and that self-consistently incorporate the contribution from
        dissipationless, monochromatic,
        hydromagnetic (with angular frequencies $\omega$ well below ion gyro-frequencies),
        toroidal Alfv\'en waves.
The waves, which smoothly pass the Alfv\'en point where the combined meridional
        Alfv\'enic Mach number $M_T$ equals 1 (cf. Eq.(\ref{eq_def_MT})),
        are coupled to the flow solely through the wave-induced ponderomotive forces.
Azimuthal symmetry about the magnetic axis is assumed throughout, and solar rotation is neglected.
However, no assumption has been made that the wavelength is small compared with the spatial scales at which the 
        solar wind parameters vary.
Starting with a waveless model, for the fast and slow solar wind alike, we obtained a grid of models corresponding to
        different $\omega$
        and base wave amplitudes $\delta u_b$,
        defined as the amplitude of the electron velocity perturbation $\left<u_e^2\right>^{1/2}$
        at 1~$R_\odot$.
The non-WKB effect is then examined, in a systematic and quantitative fashion, 
        by comparing the models with that incorporating WKB waves with the same $\delta u_b$,
        as shown by Figures~\ref{fig_fast_par} and ~\ref{fig_slow_par}.
The results may be summarized as follows:
\begin{enumerate}
 \item
The non-WKB effects are significant for the majority of the examined solutions.
In comparison with comparable WKB waves, the non-WKB ones are less effective 
      in accelerating the solar wind inside the Alfv\'en point.
As a consequence, non-WKB Alfv\'en waves may produce a proton flux up to 32\% lower,
      an Alfv\'en point 29\% further from the Sun than their WKB counterparts
      for fast solar wind solutions.
Even though the ion flux ratio increases by $\lesssim 10\%$ from non-WKB to WKB models,
      indicating that the WKB waves boost the fluxes of both ion species to a similar extent,
      the maximum of the alpha abundance, $n_{\alpha p,M}$, is considerably affected by the non-WKB effect,
      with the non-WKB values being up to 77\% higher than the WKB ones.
The differences between WKB and non-WKB models are more prominent for
      slow solar wind solutions, 
      for which the non-WKB models produce values for $n_{\alpha p, M}$
      up to $92\%$ higher.
It is also found that the influence associated with non-WKB effects tends to decrease with increasing frequency:
      the waves in the fast (slow) winds with $\omega \gtrsim 3.5\times 10^{-3}$ 
      ($1.5\times 10^{-3}$) \angfrequnits\ 
       may be described by the WKB limit with an accuracy to better than 10\%.

\item 
While the Alfv\'en waves tend to reduce the magnitude of the proton-alpha
      speed difference $|U_{\alpha p}|$ in general, different mechanisms
      operate in two different regimes, 
      separated by a critical frequency $\omega_c$,
{\revise which in principle may be different from model to model
      especially when the wave base amplitude $\delta u_b$ is large. 
      For the range in which $\delta u_b$ is varied in the
      examined numerical solutions, however, 
      $\omega_c$ hardly varies with $\delta u_b$}
      and is found to be $\sim 3\times 10^{-5}$ ($\sim 1.3\times 10^{-5}$)\angfrequnits\
      for the fast (slow) solar wind
      models.
When $\omega > \omega_c$, the fluctuations are wave-like and tend to accelerate both ion species, thereby
      losing most of their energy in doing work on ion flows.
On the other hand, when $\omega < \omega_c$, a quasi-static behavior results: the fluctuations may
      act to accelerate the slower 
      flowing ion species but decelerate the faster moving one in a large portion of the computational domain,
      and only a minor fraction of the wave energy flux injected at the base is lost.
The fluctuations with the lowest frequency are no less efficient in reducing $|U_{\alpha p}|$ than the WKB waves:
      in the slow solar wind solutions, they may be able to quench 
      a significant $|U_{\alpha p}|$ with base amplitudes as small as $4$\velunits.
\end{enumerate}

{\revise Before proceeding, we note that to examine the wave effects on the solar wind, an alternative approach will be
     to directly solve the multi-fluid equations where the Alfv\'en waves are introduced via boundary conditions
     \citep[cf.][]{Ofman_04}.
In our study, monochromatic waves are invoked so that a detailed understanding of the frequency dependence of
     the solar wind properties may be found.
On the other hand, \citet{Ofman_04} introduced a broad spectrum of Alfv\'en waves 
     into a resistive and viscous solar wind plasma.
Consequently, a comparison of his model with ours is not straightforward.
Nevertheless, in a previous study~\citep{OfmanDavila_98}, where monochromatic Alfv\'en waves were used
     and a set of isothermal single-fluid MHD equations were solved, 
     the authors concluded that their results agree with the studies
     of a single-fluid isothermal wind as conducted by \citet{MacChar_94}, who
     used essentially the same approach as ours. 
From this we may conclude that the results obtained in this paper may as well be supported by direct
     numerical solutions, the construction of which will be presented in a future publication.

Another way to extend the current model is to use a spectrum of Alfv\'en waves instead of monochromatic ones.
This will also be more consistent with observations which show that in reality, the Alfv\'enic fluctuations
     span a broad spectrum and do not possess a preferred frequency~\citep[e.g.][]{TuMarsch_95}.
With this caveat in mind, the presented study nonetheless provides a better understanding of the underlying physics.
For instance, when a spectrum is taken into account, one may expect that net effect will also be that 
     the magnitude of the proton-alpha speed difference is reduced, and the reduction
     will have some dependence on the spectral slope.
In addition,} the existence of an $\omega_c$ of the order of several times
       $10^{-5}$\angfrequnits\ may have significant consequences on the spectrum of the
       velocity perturbations of alpha particles,
       denoted by $P_\alpha$ for brevity.
In the present study we have concentrated on monochromatic waves
       for the purpose of presenting a systematic study on the effects brought forth by the finite wavelength.
However, from the results obtained in the present paper and paper I~\citep{LiLi07},
       it is possible to deduce some properties of $P_\alpha$
       in the presence of transverse Alfv\'enic fluctuations.
From an observational point of view, the spectrum of proton velocity fluctuations, $P_p$,
       is less extensively studied
       than that of the magnetic field fluctuations $P_B$,
       because of the considerably lower temporal resolution of plasma instruments
       relative to that of magnetometers \citep[cf.][]{Podesta_etal_06, Podesta_etal_07}.
Measuring $P_\alpha$ is even more difficult given that
       the alpha particles are more tenuous.
As a result, such a study on $P_\alpha$ has yet to appear but is certainly feasible for plasma instruments on board future
       missions such as Solar Orbiter or Solar Probe.
This is because these missions may approach the Sun
       as close as 4~$R_\odot$, and therefore the alpha particles to be sampled
       are much hotter and have a much larger number density than in the near-Earth regions.

Let us now picture what $P_\alpha$ in fast solar winds may look like
       in the super-Alfv\'enic regions (beyond say $\sim 40$~$R_\odot$ where the Alfv\'enic Mach number
       $M_T \gtrsim 3$), supposing that
       the waves are propagating parallel to a radial magnetic field, and that
       $U_{\alpha p} = U_A$, $x_\alpha = 0.2$ and $x_p = 0.8$,
       where $x_k = \rho_k/\rho$ ($k=p, \alpha$).
For frequencies corresponding to $\omega > \omega_c$, the fluctuations are genuinely wavelike
       and one expects that a WKB behavior results.
It then follows from Equations (25)
       and (41) in paper I that the ratio of the amplitude of alpha to proton velocity perturbations 
       $\eta \approx (1-x_p-x_p x_\alpha/2)/(2-x_p-x_p x_\alpha/2) \approx 0.11$.
On the other hand, for $\omega < \omega_c$, a quasi-static manner results,
       and an analogy is readily drawn with the problem of angular momentum transport in a multicomponent
       solar wind. 
In this case, Equations~(2a) and (2b) in \citet{Li_etal_07}, which are equivalent to the zero-frequency 
       limit treated in paper I, are more convenient to use and yield that
       $\eta \approx (\rho_p U_p)/(\rho_\alpha U_\alpha) \gtrsim 3$ by noting that
       the amplitudes are largely determined by the terms associated with $U_{\alpha p}$.
Consequently, in the spectrum of the velocity fluctuations of alpha particles $P_\alpha$, one may see
       an apparent spectral break at $\omega_c$
       around which the spectrum shows a steep slope provided that
       the proton velocity fluctuation spectrum is somehow smooth here.
Whether $P_\alpha(\omega)$ behaves like this in reality remains to be tested by future 
       in situ measurements.

\acknowledgements
This research is supported by an STFC (PPARC) rolling grant to the Aberystwyth University.
We thank the anonymous referee for the constructive comments which helped to improve this paper substantially.


\end{document}